# The physical origin of the maximum stellar size

Gautham N. Sabhahit,[1] Jorick S. Vink,[1]

[1]*Armagh Observatory and Planetarium, College Hill, Armagh, BT61 9DG, N. Ireland, UK*

**One might expect the most massive stars to also be the largest by size, yet they are not, and this has puzzled astronomers for decades. The Eddington limit sets an upper bound to stellar mass through the balance between radiation pressure and gravity[1], but a second empirical boundary on stellar radius has equally far-reaching consequences for modern Astrophysics: it affects the black-hole masses and merger rates inferred by LIGO and Virgo[2], determines whether a star remains a hot, compact ionising source or inflates into a cool supergiant, and thereby shapes how spectra of distant, unresolved stellar populations are interpreted. Half a century ago, this radius limit was shown to trace a characteristic kinked shape in the Hertzsprung–Russell diagram known as the Humphreys-Davidson limit[3], yet a predictive, first-principle physical explanation has remained elusive. Here we show that this kink is the evolutionary manifestation of a transition from classical stellar outflows to an Eddington-enhanced mass-loss regime implemented self consistently in evolutionary models. With our models accurately reproducing the empirical constraints on the radii of the most massive stars, we now have a framework that can be incorporated into binary population synthesis, black-hole mass predictions, gravitational-wave event rates, and the interpretation of high-redshift James Webb spectra[4].**

Almost 50 years ago, Humphreys and Davidson[3] published their upper-luminosity limit to stars. Although this empirical HD limit was soon linked[5] to the theoretical radiation-pressure limit derived by Arthur Eddington, a convincing physical explanation for its characteristic kinked shape (Fig. 1) has yet to be provided.

For decades, astronomers pondered on the question as to why stars expand after core-hydrogen burning to become red giants and, while this is important for stellar evolution, and even civilisations whose planets might one day be engulfed by them, the Universe itself remains rather unaffected. For massive stars it is the reverse question - why the most massive stars do *not* become red supergiants (RSGs) - that is astronomically decisive, far beyond stellar evolution alone.

Contrary to common perception, the HD-limit is not primarily a limit to stellar luminosity or mass, but a limit to stellar radii. This distinction is crucial, as stellar size determines whether stars interact with companions and whether they remain compact, hot ionising sources or instead evolve into cool supergiants that can reach as large as thousands of $R_\odot$ (Fig. 1). The stellar radius ultimately decides the evolution and fate of the most massive stars as supernovae or black holes[6,7], bifurcating whether they keep their H envelopes intact producing H-rich Type II SNe or stripped-envelope SNe of Type Ibc[8]. It is also a key parameter for massive binary evolution, which is relevant as the majority of massive O-type stars are part of a close binary system[9], and stellar radius affects the interaction probability[10] - influencing Roche-lobe overflow, common-envelope evolution, and stellar merging, thereby affecting the entire field of high-energy astrophysics, including X-ray binaries and gravitational-wave astronomy with interferometers such as LIGO/VIRGO[2].

The fact that size matters is equally profound for the radiative and chemical evolution of galaxies. Stars that remain compact continue to emit copious amounts of ionising radiation, while stars with RSG dimensions contribute no ionising fluxes to their surrounding galaxies at all. The mass-loss physics that sets the HD-limit prevents stars from radius enlargement, thus allowing the most massive stars to become a key contributor for the chemical and radiative feedback of the Early Universe[11,12] with the JWST[4] and may even have caused cosmic reionisation[13].

The same decade of the 1970s witnessed the birth of the radiation-driven wind theory by Lucy and Solomon[14] and Castor et al.[15] (CAK model), where it was shown that radiation pressure on spectral lines in an outflowing wind could remove mass from stars. These CAK-type models were able to account for the UV P-Cygni lines in normal OB supergiant spectra observed with satellites such as IUE and HST. These wind investigations culminated in widely used OB-star mass-loss predictions from Monte Carlo (MC) radiative transfer models[16]. However, the predicted O-star mass-loss rates ($\dot{M}$) from these classical models fall well short of what would be required to prevent the most massive stars from inflating[17,18] and explain the HD-limit.

A decade ago, two key papers introduced the concept of the transition mass-loss rate[19,20]. Here, optically-thin outflows undergo a phase transition when multiple line scattering becomes important, causing an upturn or "kink" in the slope of the mass outflow rate with Eddington parameter $\Gamma$ (Fig. 2), which goes as the luminosity-to-mass ($L_* / M_*$) ratio. The known very massive stars (VMSs) with $M > 100\ M_\odot$ in the Local Universe indeed show strong Wolf-Rayet type emission wind features[21,22,23], and evidence of VMS wind signature has also been found at higher redshift[24]. While the existence of a $\dot{M}$–$\Gamma$ kink had been recognised[19] and observationally confirmed[25], no quantitative description of a kink-bearing mass-loss rates were available that could be implemented self-consistently in stellar evolution calculations. More recently, hydrodynamically consistent atmosphere models have made this possible[26], predicting mass-loss rates directly from the wind dynamics and naturally recover the transition to the high-$\Gamma$ regime.

In Fig. 3, we show that when this kink-bearing mass-loss rates is implemented self-consistently into MESA (Modules for Experiments in Stellar Astrophysics) stellar evolution calculations, a corresponding characteristic kink in the HD-limit naturally emerges. VMSs above 100 $M_{\odot}$ are already close to their Eddington limit at birth, and so enter the high-$\Gamma$ mass-loss regime almost immediately (as shown by how early in their MS lifetime the mass-loss kink occurs in Extended Data Figure 1), lose sufficient envelope mass to suppress inflation, remaining at high effective temperatures and continuing to provide copious ionising radiation.

Stars in the intermediate 30–100 $M_{\odot}$ range begin their lives with more modest $\Gamma$ values and only transition into the high mass-loss regime late in their evolution. Correspondingly, as the initial mass decreases, these stars undertake longer excursions toward cooler temperatures before strong mass loss sets in and they avoid becoming RSGs. It is this delayed onset of strong mass loss that sets the slanted segment of the HD-limit.

However, this slant trend does not continue indefinitely; below $\log(L/L_{\odot}) \sim 5.5$, the boundary flattens into a luminosity ceiling instead[27,28]. This is because stars below ~30 $M_{\odot}$ do not encounter the mass-loss kink at all. With only modest mass-loss rates throughout their lives, these likely Type II SN progenitors retain most of their H-rich envelope and expand unimpeded to RSG proportions, reproducing the luminosity ceiling.

The shape of the HD-limit, including its characteristic kink, thus naturally emerges as the evolutionary manifestation of the kink in the $\dot{M}$–$\Gamma$ relation as predicted by radiation-driven wind theory. For the first time, we provide a predictive, self-consistent framework that recovers the HD-limit from first principles, a step beyond current state-of-the-art models which have investigated the limit by invoking the empirical boundary itself (e.g., Pauli et al.[29]).

Having identified the physical mechanism that reproduces the HD-limit, robust predictions can now be made across many differing areas of Astrophysics. Our predictions can by utilised by the community of binary population synthesis[30] directly governing whether massive binaries interact through Roche-lobe overflow, common-envelope evolution, stellar mergers, and so critical for making reliable predictions of gravitational wave events rates detected by LIGO/Virgo[2].

Accurate reproduction of the HD-limit also means that the ionising photon output of VMS populations can now be reliably predicted. This will enable self-consistent inclusion of VMS contributions in *spectral* population synthesis models (e.g., Starburst99[31]) and a better understanding of their role in chemical pollution of globular clusters[32,33,34], star-forming clusters and galaxies[24,35], especially given the new window on high-redshift stellar populations opened up by JWST[4,36].

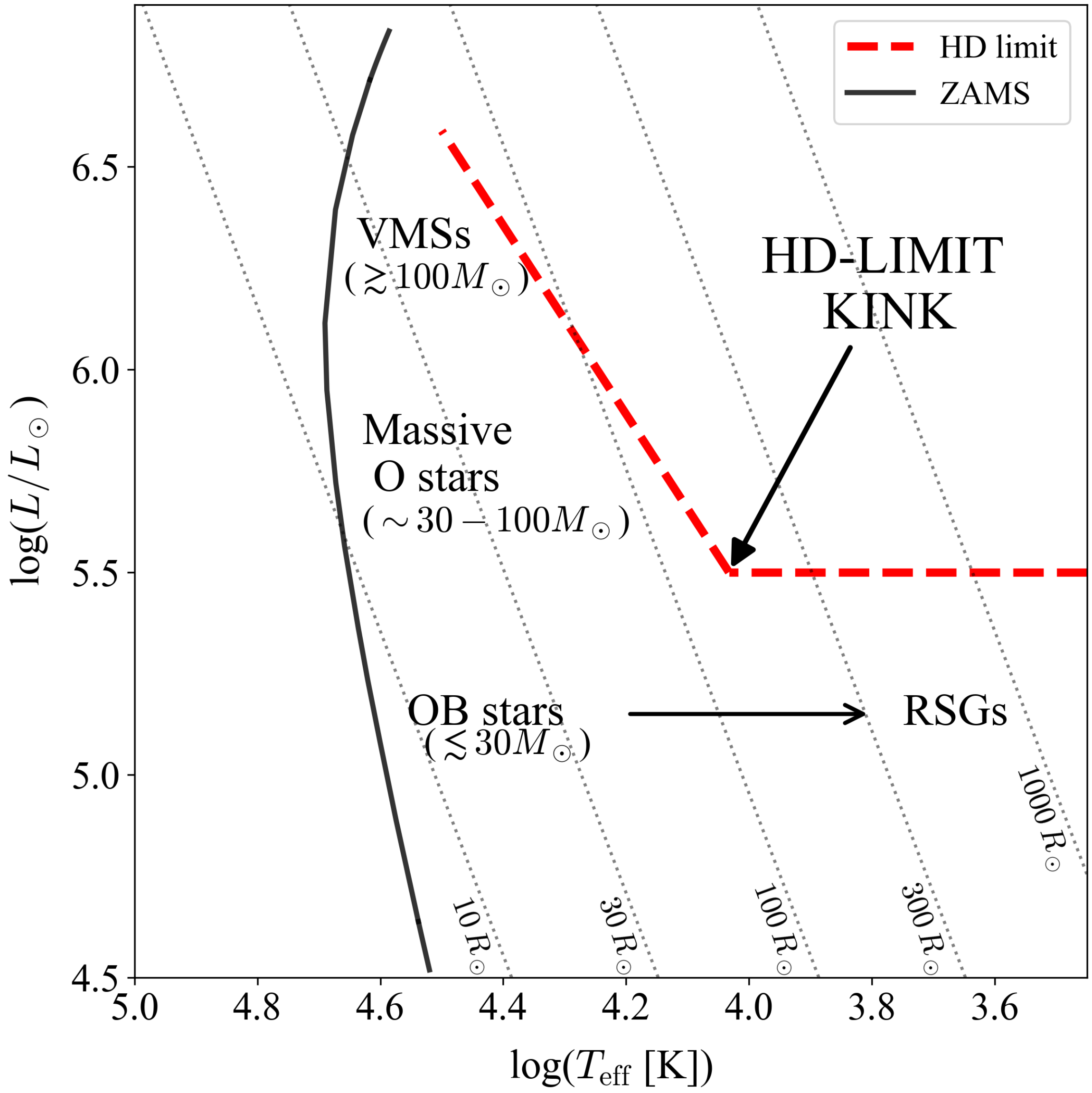

**Figure 1.** Hertzsprung-Russell diagram showing the empirical HD-limit kink (dashed red)[1,29,30]. Constant radius lines (dotted black) are overplotted showcasing the fundamental radius limit imposed by the HD-limit. The location of the Zero Age Main Sequence is shown for comparison (solid black).

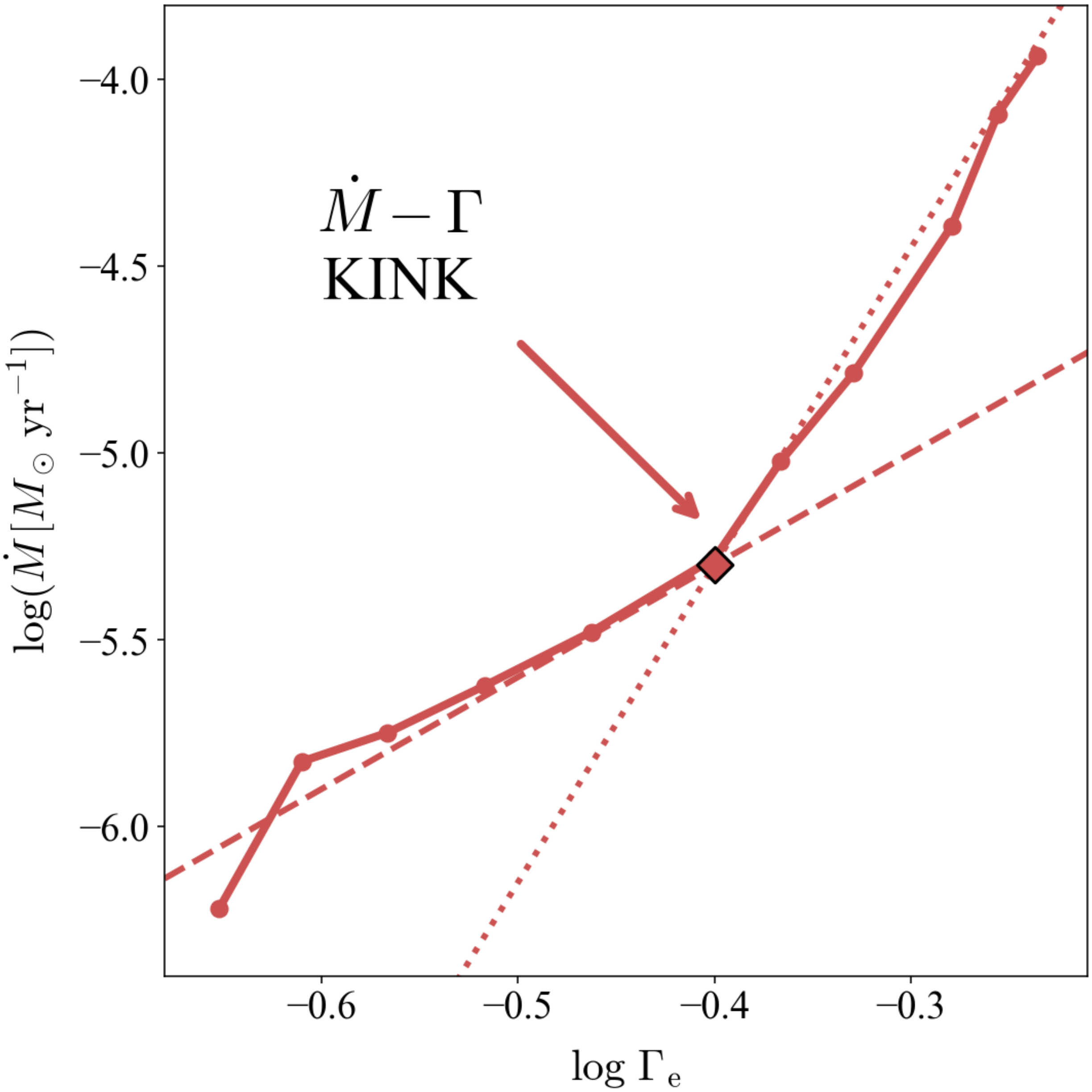

**Figure 2.** Predicted mass-loss rates from PoWR$^{HD}$ models as a function of electron scattering Eddington parameter illustrating the kink in the $\dot{M}$-$\Gamma$ scaling.

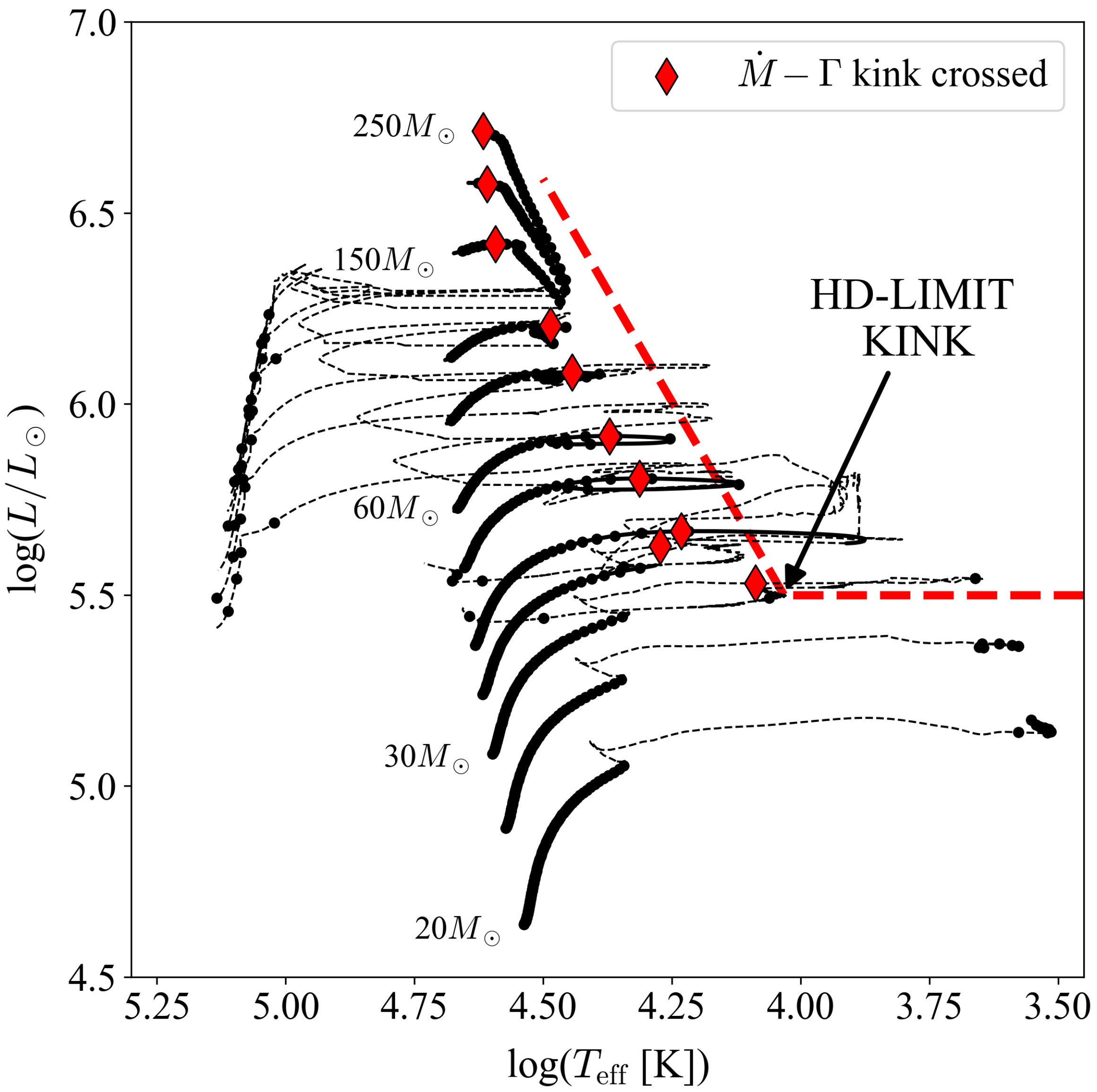

**Figure 3.** Hertzsprung-Russell diagram tracks of our evolutionary models with dot symbols mark timesteps of every 100,000 years. Red diamond symbols mark the point at which each model crosses the mass-loss kink, above which $\dot{\rm M}$ scales steeply with the Eddington parameter.

**Methods**

We used the Modules for Experiments in Stellar Astrophysics (MESA, v. 12115)[37, 38, 39, 40, 41] code to compute the stellar structure and evolutionary tracks presented in this work. Models were evolved from the pre-main sequence through to the end of core-He burning for initial masses of 20, 25, 30, 35, 40, 50, 60, 80, 100, 150, 200, and 250 $M_{\odot}$. The initial metal mass fraction was set to Z = 0.02 with individual metal abundances following Grevesse and Sauval[42]. The helium mass fraction was calculated as $Y = Y_{prim} + (\Delta Y/\Delta Z) Z$, with $Y_{prim} = 0.24$ and $\Delta Y/\Delta Z = 2$, and the H mass fraction follows from $X + Y + Z = 1$.

Convection was treated using mixing length theory[43] with the free parameter $\alpha_{MLT} = 1.5$. The MLT++ routine of MESA was switched off during the MS, meaning we use the temperature gradient as predicted by MLT without artificially reducing it in super-adiabatic layers. It was switched on for post-MS evolution to ensure numerical stability during core-He burning. Convective stability was assessed using the Ledoux criterion. Semiconvective mixing in regions of chemical gradient was implemented using the diffusive scheme of Langer et al.[44] with an efficiency of $\alpha_{sc} = 1$. Convective boundary mixing was treated with overshooting at the core boundary using an exponential formalism from Herwig[45] with $f_{ov} = 0.03$. All models were initially rotating at 20% of their critical rotation rate.

We used the basic.net with eight isotopes for the reaction network. Opacities were taken from the OPAL Type 2 tables, which account for variations in the mass fractions of metals, mainly of carbon and oxygen mass fractions. The equation of state followed the OPAL EOS tables[46] blended with additional tables[37].

The novelty of the present work is the self-consistent implementation of mass-loss rates as predicted directly from hydrodynamically consistent wind-atmosphere models into stellar evolution calculations. In a series of works[26,47], we investigated the wind properties of massive and very massive stars at $Z = 0.02$ using the hydrodynamic version of the Potsdam Wolf-Rayet wind-atmosphere code (PoWR$^{HD}$), which was updated[48,49] to account for wind hydrodynamics self-consistently as part of the usual iterative process of solving radiative transfer in an expanding atmosphere, non-LTE statistical rate equations, and the radiative equilibrium equation. The PoWR$^{HD}$ code solves the wind hydrodynamics and allows mass-loss rates and wind velocity stratifications to be predicted from first principles, accounting for the various forces acting in the atmosphere, rather than being prescribed as inputs. We provided mass-loss rate predictions applicable to the 20-500 $M_\odot$ mass range, which takes as input relevant stellar parameters such as the Eddington parameter $\Gamma_e$, luminosity $L_*$, and temperature $T_{eff}$ and outputs the mass-loss rate (supplementary material).

The hydrodynamically consistent mass-loss rates from the PoWR$^{HD}$ code are used for surface effective temperatures in the range $3.8 < \log(T_{eff}/K) < 4.9$. Below $\log(T_{eff}/K) = 3.7$, we adopt the standard de Jager RSG mass-loss rates[50], and above $\log(T_{eff}/K) = 5$, we use the hydrodynamically consistent classical Wolf-Rayet predictions also from the PoWR$^{HD}$ code[51]. Between these regimes, the mass-loss rates are interpolated smoothly between the adjacent regimes.

## Extended data items

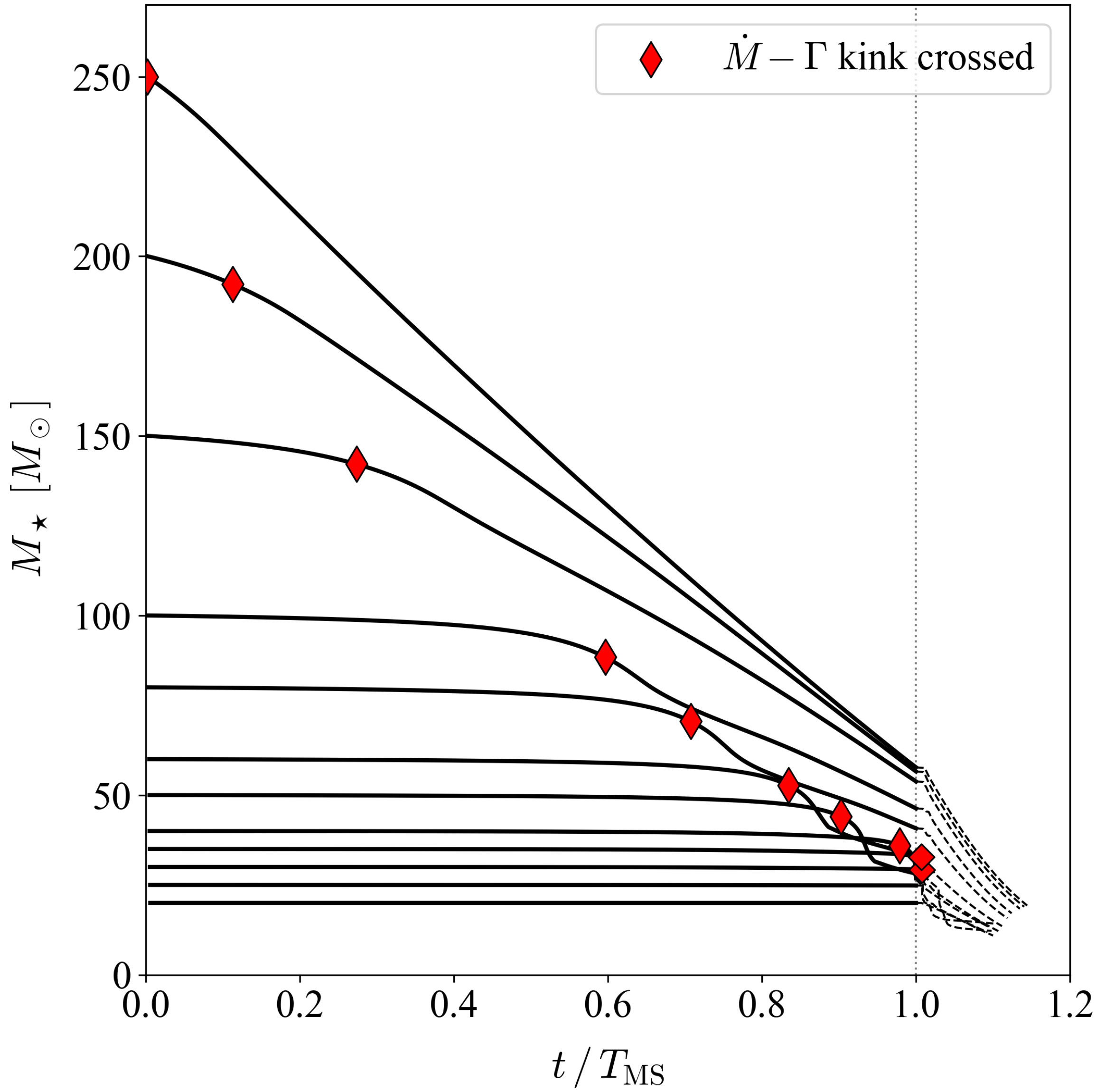

**Extended Data Figure 1.** Total mass evolution plotted against age normalised to the main-sequence lifetime. Red diamond symbols mark the point at which each model crosses the mass-loss kink, above which $\dot{\mathrm{M}}$ scales steeply with the Eddington parameter.

**Correspondence and requests for materials** should be addressed to Gautham N. Sabhahit (gauthamns96@gmail.com) and Jorick S. Vink (jorick.vink@armagh.ac.uk).

## Acknowledgements

JSV and GNS are supported by the STFC grant ST/Y001338/1 (PI Vink).

## Author contributions

GNS performed the numerical simulations. GNS and JSV wrote the Manuscript.

## Competing interest declaration

The authors declare no competing interests.

## Data availability

The evolutionary model outputs (MESA tracks) will be made available from the corresponding author upon reasonable request.

## Code availability

The models presented in this work utilise the publicly available stellar structure and evolution code Modules for Experiments in Stellar Astrophysics (MESA, v. 12115). The exact version of the code is available (https://zenodo.org/records/3473377)

**Supplementary Information**

The hydrodynamically consistent $PoWR^{HD}$ models predict a pronounced kink in the mass-loss rate versus Eddington parameter relation (Fig. 2). The resulting prescription is primarily controlled by the classical Eddington parameter $\Gamma_e$ and reproduces the transition between the classical O-star regime and the enhanced mass-loss regime associated with multiple scattering, where

$$\Gamma_e = \frac{\kappa_e L_*}{4\pi G c M_*}$$

with $k_e$ the Thomson opacity, $L_*$ the luminosity, and $M_*$ the stellar mass. At temperatures characteristic of O stars and VMSs, hydrogen and helium are fully ionised, keeping $k_e$ essentially fixed and therefore $\Gamma_e$ hardly changes with depth and provides a relatively constant measure of the model's proximity to its radiative limit. Additional dependences on luminosity, temperature and surface composition are included as described in Sabhahit et al.[26]

We anchor our mass-loss predictions to independent empirical constraints. First, our rates reproduce the transition mass-loss rate of Vink and Gräfener[20] applied to the Arches cluster. If the observed Of/WNh stars[52] indeed mark the transition between optically thin and optically thick winds, a robust mass-loss estimate follows directly from momentum arguments: $\dot{M}_{trans} \approx 0.6\, L_*/(v_\infty c)$. This transition mass-loss rate serves as a robust anchoring point. What is otherwise a complex, model-dependent mass-loss empirical determination based on Hα-recombination diagnostics reduces to a mere luminosity estimate, and is therefore insensitive to model assumptions, for example of clumping. For the Arches Cluster, the transition mass-loss rate yields $\log(\dot{M}_{trans}\ [M_\odot\ yr^{-1}]) \approx -5.2$, in close agreement with our value of $\log(\dot{M}_{PoWR\text{-}HD}\ [M_\odot\ yr^{-1}]) \approx -5.15 \pm 0.1$ where the optical depth in the wind crosses unity[27].

We also compare our rates to empirical $\dot{M}$–$\Gamma_e$ compilation of Pauli et al.[53], who use accurate dynamical masses where available and stars with agreeable spectroscopic and evolutionary masses for $\Gamma_e$ calculations (Supplementary Fig. 1). Our rates agree well with the empirical compilation within the scatter as expected since our predictions are predominantly set by $\Gamma_{e.}$

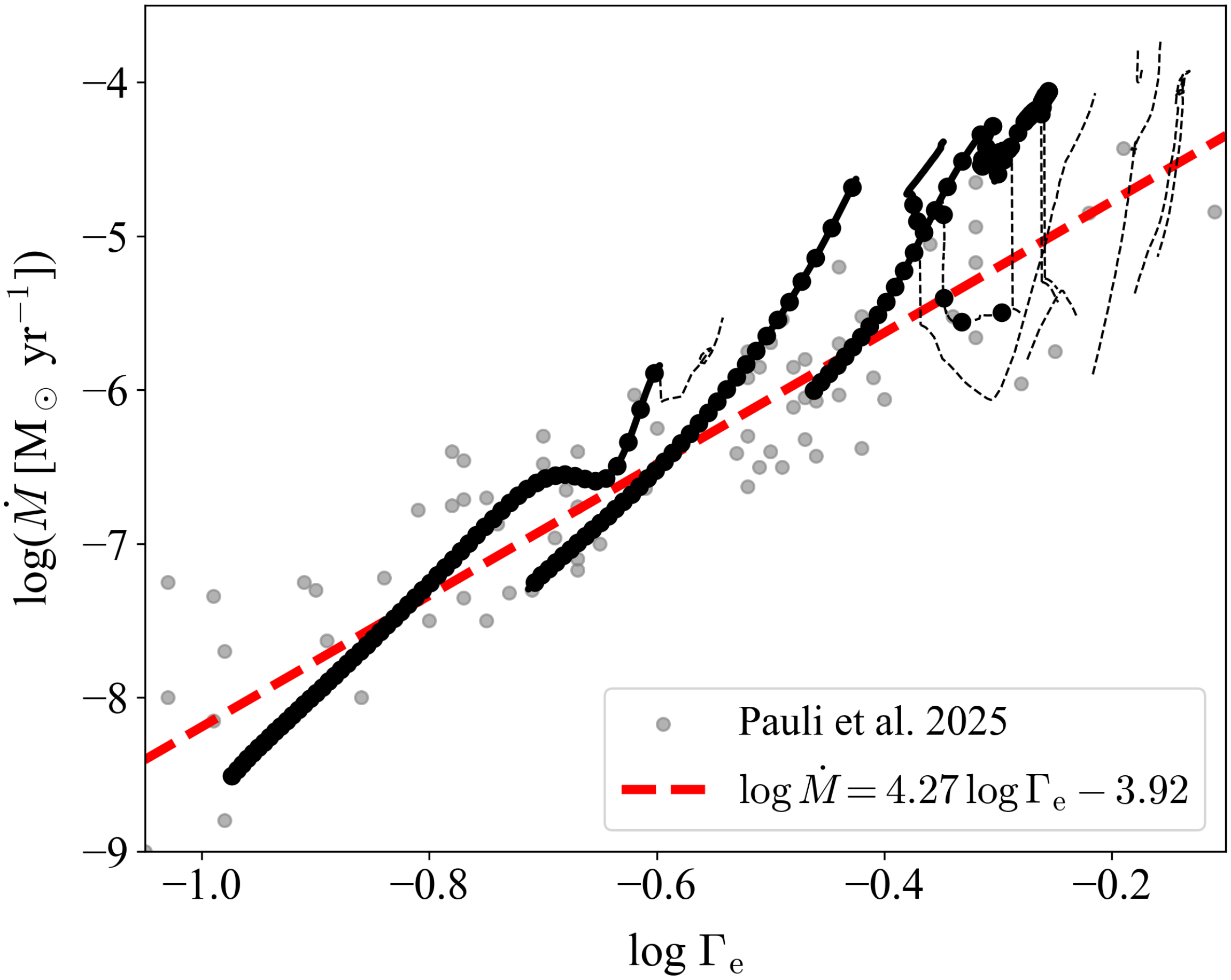

**Supplementary Figure 1.** Mass-loss rate as a function of $\Gamma_e$ when implemented self-consistently within stellar evolutionary models. The grey circles are empirical data[53], and the dashed red line is best-fit line to the observations.